\documentclass[pra,a4paper,twocolumn,superscriptaddress,longbibliography,nofootinbib]{revtex4-2}

\usepackage[dvipsnames]{xcolor}
\usepackage{amsmath,amssymb}
\usepackage{bm}
\usepackage{graphicx}
\usepackage[colorlinks]{hyperref}

\begin{document}

\title{Manipulation of magnetic skyrmions by 
non-uniform electric fields}

\author{N. I. Simchuck}

\affiliation{Laboratory for Condensed Matter Physics, HSE University, 101000 Moscow, Russia}

\author{I. S. Burmistrov}

\affiliation{\mbox{L. D. Landau Institute for Theoretical Physics, Semenova 1-a, 142432, Chernogolovka, Russia}}

\author{S. S. Apostoloff}

\affiliation{\mbox{L. D. Landau Institute for Theoretical Physics, Semenova 1-a, 142432, Chernogolovka, Russia}}

\affiliation{Laboratory for Condensed Matter Physics, HSE University, 101000 Moscow, Russia}

\begin{abstract}
Magnetic skyrmions are topologically protected spin textures in ferromagnetic materials that hold great promise for both classical information storage and processing, as well as for fault-tolerant quantum computing. Realizing practical skyrmion-based devices demands an energy-efficient and precise method for their flexible manipulation. In this paper, we theoretically propose such a tool, leveraging the magnetoelectric effect induced by a localized electric field generated by one or several charged tips. Combining complementary numerical simulations and analytical approaches, we develop a consistent theory describing the stability and dynamics of Néel-type skyrmions under the influence of the electric field from a charged tip. Specifically, we demonstrate that the electric field can create, drive, and annihilate skyrmions of both chiralities, as well as more complex textures such as skyrmioniums and target skyrmions. We identify several distinct dynamical regimes of skyrmion motion near the tip and map them onto a phase diagram. Finally, we discuss the feasibility of a practical device capable of controlled skyrmion manipulation based on this principle.

\end{abstract}

\maketitle

\section{Introduction} 

Magnetic skyrmions---whirl-like spin textures in chiral ferromagnetic materials---were theoretically predicted in the 1980s~\cite{Bogdanov1989} and experimentally observed only recently~\cite{Muhlbauer2009,Seki2012}. Owing to their topological nature~\cite{Nagaosa2013}, skyrmions are extraordinarily metastable and exhibit particle-like behavior, offering new opportunities in spintronics and in information storage and processing~\cite{Back2020}. These prospects motivate the development of effective tools for skyrmion manipulation, including their creation, motion, resizing, and annihilation. Current approaches~\cite{Litzius2020} typically employ electric currents in the ferromagnetic layer or in an adjacent metallic substrate, which interact with the topologically nontrivial magnetization distribution to drive skyrmion motion. However, several drawbacks---such as Joule heating~\cite{Brock2020}, the mismatch between the directions of skyrmion motion and current flow caused by the skyrmion Hall effect~\cite{Litzius2017}, and the inherently nonlocal action of the currents---highlight the need for more flexible and versatile control methods. Among the various alternatives, external nonuniform magnetic~\cite{Shustin2023,Apostoloff2024} and electric~\cite{Pyatakov2015,Wang2019,Wang2022} fields are of particular interest because they can be applied locally and used to manipulate, create, or annihilate individual skyrmions.

The interaction between the skyrmion magnetization and the external magnetic field is straightforward: the spins tend to align with the field, minimizing the Zeeman energy. A uniform magnetic field determines the equilibrium skyrmion size, whereas a nonuniform field additionally influences its position and shape~\cite{Andriyakhina2022,Apostoloff2024}. Hence, in principle, magnetic fields can be employed to manipulate skyrmions. A notable example of a nonuniform magnetic-field source is a superconducting vortex in a substrate layer. The skyrmion and the vortex can attract each other to form a stable bound pair---either coaxial or eccentric---depending on material parameters~\cite{Menezes2019,Petrovic2021,Xie2024,Apostoloff2025}. The coaxial configuration is of particular interest because theoretical predictions~\cite{Yang2016,Rex2019} suggest that such hybrid topological structures may host Majorana electronic states, providing a possible platform for topological quantum computations~\cite{Nothhelfer2022,Konakanchi2023}.
A key experimental challenge lies in generating magnetic fields localized on the characteristic skyrmion length scale. 

In contrast, nonuniform electric fields can be realized more easily, for instance by applying voltage to a thin metal tip~\cite{Pyatakov2015,Wang2022}. Due to the magnetoelectric effect (ME)~\cite{Yablonsky1983,Mostovoy2006,Dzyaloshinskii2008}, a nonuniform magnetization induces an electric polarization,  cf. Eq.~\eqref{eq:Pm}, and the applied electric field interacts with this polarization to drive domain walls and skyrmions~\cite{Pyatakov2015-ufn,Wang2019}. In conducting thin films, an applied electric field also partially penetrates the material, inducing local charge inhomogeneities that further influence the nonuniform magnetization. To avoid additional complications, we focus here on insulating magnetic materials, anticipating that the qualitative conclusions remain valid for conducting systems as well.

\begin{figure}[b]
	\centering
	\includegraphics[width=1\linewidth]{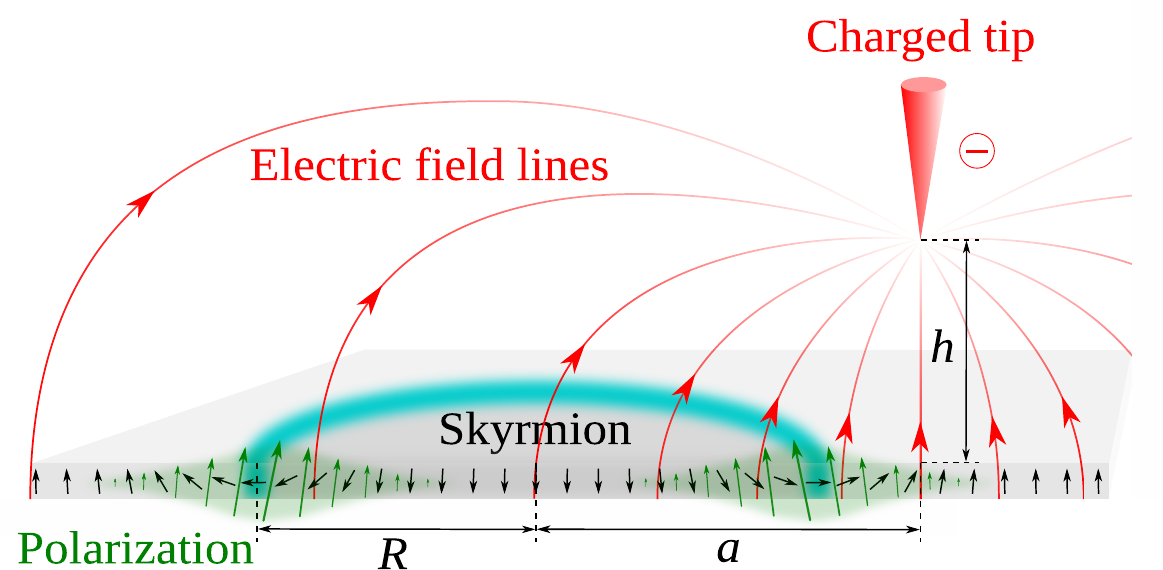}
	\caption{ Scheme of the setup. The skyrmion in the thin ferromagnetic film subdued by the electric field from the charged tip. The black arrows correspond to the magnetization, while green arrows with shadings illustrate the ME-induced polarization.
		 }
	\label{fig:scheme}
\end{figure}

In this work, we develop a theory for a metastable N\'eel-type skyrmion states near a point-like source of electric field (see Fig.~\ref{fig:scheme}). Using an analytical approach based on the symmetric skyrmion profile, cf. Eq.~\eqref{eq:m_rad_sym}, and 
confirming 
it with micromagnetic simulations, we demonstrate that an electric field can not only move and resize existing skyrmions but also create and annihilate them. Furthermore, we show that the electric field can stabilize skyrmions of both chiralities as well as so-called target skyrmions. Finally, we propose a possible realization of a device for electrically controlling skyrmions in a ferromagnetic film (see Fig.~\ref{fig:traject}).

The paper is organized as follows. In Sec.~\ref{sec:model} we present a detailed description of the model and methods employed in our theoretical study of skyrmion dynamics in a chiral ferromagnetic film subjected to an electric field. In particular, we discuss the magnetoelectric effect that drives skyrmion motion, the Landau–Lifshitz–Gilbert and Thiele equations governing magnetization dynamics, and the spatial profile of the electric field generated by a single charged tip.
In Sec.~\ref{sec:states}, we analytically characterize the possible (quasi-)stable configurations of a skyrmion located at a distance $a$ from the tip. Interestingly, while truly stable skyrmion states are coaxial with the tip ($a = 0$), the analysis of quasi-stable states for $a \neq 0$ provides valuable qualitative (and in some cases quantitative) insight into the dynamics of a moving skyrmion as the distance $a$ varies. We conclude this section with the phase diagram shown in Fig.~\ref{fig:diagram}, which maps the regions of possible skyrmion quasi-stable states in the parameter plane of the skyrmion–tip distance and the electric field intensity.
The skyrmion dynamics using both analytical approaches based on the Thiele equation and numerical micromagnetic simulations are studied in Sec.~\ref{sec:motion}. We demonstrate that skyrmions can be created and annihilated by switching the voltage applied to the tip.
In Sec.~\ref{sec:platform} we introduce a conceptual device architecture comprising multiple charged tips capable of manipulating skyrmions in a ferromagnetic film.
Finally,
we summarize our findings and outline potential directions for future research and extensions of the present work in Sec.~\ref{sec:discuss}.

\section{Model and methods\label{sec:model}}

We consider a thin chiral ferromagnetic film whose magnetic free energy is given by~\cite{Bogdanov1989}:
\begin{equation} 
	\frac{\mathcal{F}_{\text{magn}}[\bm{m}]}{d_F} 
	= \int d^2 \bm{r} \bigg\{ A (\nabla \bm{m})^2 + K(1- m_z^2) + w_{\rm DM}[\bm{m}] \bigg\}, 
	\label{eq:MagFe} 
\end{equation}
where $\bm{m}(\bm{r})$ denotes a unit vector in the direction of magnetization, $d_F$ is the film thickness. The magnetic free energy~\eqref{eq:MagFe} is normalized such that $\mathcal{F}_{\text{magn}} = 0$ for the ferromagnetic state with $m_z = 1$ where the $z$~axis is oriented normal to the film interface. The parameters $A > 0$ and $K > 0$ are the exchange stiffness and effective\footnote{The contribution of the demagnetizing field is included in the effective perpendicular anisotropy constant, $K = K_0 - 2\pi M_s^2$.~\protect\cite{Menezes2019,Andriyakhina2021,Kuznetsov2022}} perpendicular anisotropy constants, respectively. The density $w_{\rm DM}[\bm{m}]$ represents the contribution from the relativistic antisymmetric exchange, also known as the Dzyaloshinskii–Moriya interaction (DMI). It depends on the crystal symmetry class of the material and gives rise to various noncollinear spin textures. Here, we focus on materials with ``pyramidal'' symmetry groups $C_{nv}$, for which the DMI contribution takes the form
\begin{equation}
	w_{\rm DM}
	[\bm{m}] = D \bigl [m_z \nabla \cdot \bm{m} - (\bm{m}\cdot \nabla) m_z \bigr ], \label{eq:w_cnv}
\end{equation}
where $D$ is the DMI constant. Expressions for $w_{\rm DM}[\bm{m}]$ corresponding to other symmetry classes can be found in Ref.~\cite{Bogdanov1989}. 
In the present study, for the sake of concreteness, we focus on the case of $D>0$, whereas most results can be easily extended to $D<0$.

\subsubsection{Magnetoelectric effect}

A nonuniform magnetization in the ferromagnetic film induces a ferroelectric polarization~\cite{Yablonsky1983,Mostovoy2006,Dzyaloshinskii2008},
\begin{equation}
	\bm{P}=\gamma_{\rm ME} \bigl [\bm{m} \nabla \cdot \bm{m} - (\bm{m}\cdot \nabla) \bm{m} \bigr ],
	\label{eq:Pm}
\end{equation}
where $\gamma_{\rm ME}$ is the magnetoelectric coupling constant. Consequently, the energy of the ferromagnetic film acquires an additional contribution with the density $-\bm{P}\cdot\bm{E}$ when an external electric field~$\bm{E}$ is applied.

In this work, we explore the use of a localized, non-uniform electric field to manipulate skyrmions. Such a field can be generated by an electrically charged thin tip~\cite{Pyatakov2015} positioned at some height~$h$ above the ferromagnetic film, see Fig.~\ref{fig:scheme}. 
Taking into account that ferromagnets, such as multiferroics or ferrites, typically exhibit a high permittivity~\cite{Spaldin2005,Pullar2012}, ranging from 10 to $10^4$, we can assume that the electric field near the film interface is oriented nearly perpendicular to the surface, $\bm{E}|_{z=+0}\approx E(\bm{r}) \bm{e}_z$. Under these conditions, the magnetoelectric energy term takes the same form as the DMI energy, cf. Eq.~\eqref{eq:w_cnv},
\begin{equation} 
	\frac{\mathcal{F}_{\text{ME}}[\bm{m}]}{d_F} = \int d^2 \bm{r} D_{\rm ME}(\bm{r}) \bigl [m_z \nabla \cdot \bm{m} - (\bm{m}\cdot \nabla) m_z \bigr ], \label{eq:F_ME_DMI} 
\end{equation} 
where $D_{\rm ME}(\bm{r})=-\gamma_{\rm ME}E(\bm{r})$ can be regarded as a spatially nonuniform DMI coefficient.

\subsubsection{Skyrmions}

The DMI in the form~\eqref{eq:w_cnv} allows the stabilization of N\'eel-type skyrmions in the ferromagnetic film~\cite{Bogdanov1989}. As follows from Eq.~\eqref{eq:F_ME_DMI}, a uniform external electric field produces the same effect as the common DMI. This can be used to control the parameters of skyrmions across the entire film~\cite{Nagaosa2013,Ruff2015}, but not their positions. If the external field is nonuniform, the skyrmion can move and change its size during motion. 

To develop an analytical approach we assume that the skyrmion, both in its stable state and during evolution, retains a radially symmetric shape that can be described by the following magnetization profile:
  \begin{equation}
    \label{eq:m_rad_sym}
    \bm{m} = \bm{e}_z \cos \theta (r)+
    \bm{e}_{r}\sin \theta (r).
  \end{equation}
Here $\bm{e}_{r}$ and $\bm{e}_{z}$ are unit vectors in the radial and axial directions of the cylindrical coordinate system with the origin at the skyrmion center, and $\theta(r)$ is commonly referred to as the skyrmion angle. This assumption enables a series of analytical results highlighting various quasistable states and evolution regimes, such as motion, creation, and annihilation. Its accuracy is verified by numerical micromagnetic simulations, see~\ref{sec:motion}. 
A more general expression for a radially symmetric shape including skyrmion helicity~$\eta$ is discussed in Appendix~\ref{app:Thiele} and is given by Eq.~\eqref{eq:m_radial_helicity}.

To determine the metastable state of a skyrmion in an external electric field~$E(\bm{r})$, one should minimize the total free energy, $\mathcal{F}=\mathcal{F}_{\text{magn}}+\mathcal{F}_{\text{ME}}$. Assuming the skyrmion center is located at a certain point~$\bm{a}$, we shift the coordinate origin to that point. The expressions for the free energy defined by Eqs.~\eqref{eq:MagFe} and~\eqref{eq:F_ME_DMI} remain unchanged except for the translation $D_{\rm ME}(\bm{r}) \to D_{\rm ME}(\bm{r}+\bm{a})$. We then substitute the radially symmetric magnetization profile from Eq.~\eqref{eq:m_rad_sym} into the total free energy, integrate over the azimuthal angle~$\phi$ (which does not enter the skyrmion profile), and, upon minimization, derive the Euler--Lagrange (EL) equation for the skyrmion angle~$\theta(r)$:
\begin{equation}
	\label{eq:euler_lagrange}
	-\cfrac{\ell_w^2}{r}\partial_r(r \partial_r \theta) + \cfrac{r^2 + \ell_w^2}{2r^2} \sin 2\theta -\cfrac{2\epsilon_{\rm eff}}{r/\ell_w}\sin^2\theta =  \ell_w\partial_r \epsilon_{\rm eff}.
\end{equation}
Here $\ell_w=\sqrt{A/K}$ is the characteristic magnetic length, known as the domain-wall width, and $\epsilon_{\rm eff}(r)=\epsilon_0 + \bar{\epsilon}_{\bm{a}}(r)$ is the effective dimensionless DMI function composed of a constant term, ${\epsilon_0=D/2\sqrt{A K}}$, and a nonuniform term,
\begin{equation}
	\bar{\epsilon}_{\bm{a}}(r)=-\dfrac{\gamma_{\rm ME}}{4\pi\sqrt{A K}}\int_{-\pi}^\pi d\phi \,E(\bm{r}+\bm{a}).
\end{equation}

Equation~\eqref{eq:euler_lagrange} should be supplemented by appropriate boundary conditions. Since the magnetization far from the skyrmion is uniform, $\vec{m}(r{\to}\infty)=\vec{e}_z$, we set $\theta(r{\to}\infty)=0$.

The other boundary condition, ${\theta(r=0)=\nu\pi}$, where $\nu$ is an integer, specifies the magnetization direction at the skyrmion center. In particular, odd $\nu$ corresponds to magnetization in the center inverted with respect to the surroundings, $\vec{m}(r=0)=-\vec{e}_z$, while even $\nu$ indicates alignment, $\vec{m}(r=0)=\vec{m}(r\to\infty)$. It should be emphasized that the absolute value~$|\nu|$ and its sign 
is equal to
the number of skyrmionic walls and their chirality\footnote{We define a skyrmionic wall at $r=r_0$ with chirality $\chi=+$ or $\chi=-$ when $m_z(r_0)=0$ and $\partial_r\theta(r_0)>0$ or ${\partial_r\theta(r_0)<0}$, respectively.}, respectively, 
see details in Appendix~\ref{sec:multi}. In the main text, we consider only states with $|\nu|\leq2$, 
since these configurations are found to have the lowest energies and are accessible in micromagnetic simulations for experimentally reasonable intensities of the electric field. Accordingly, we refer to the solutions with $\nu=0$, $|\nu|=1$, and $|\nu|=2$ as the non-skyrmion, skyrmion, and skyrmionium (or $2\pi$-skyrmion) states, respectively. Note that skyrmions with $|\nu|\geq2$ are also known as target skyrmions due to their specific spin texture with alternating concentric circular domains~\cite{Tretiakov2021}.

The topological protection of a skyrmion is related to the conservation of the so-called topological charge~$Q$ that in 2D takes the following form,
\begin{eqnarray}\label{eq:rho_Q}
  	Q = \int d^2\bm{r}
    \,\rho(\bm{r}),
  	\quad	
  	\rho(\bm{r}) = -\dfrac{\bm{m}}{4\pi} 
     \cdot \left[\frac{\partial {\bm{m}}}{\partial x} \times \frac{\partial {\bm{m}}}{\partial y}\right].
\end{eqnarray}
Calculating~$Q$ for the skyrmion configuration given by Eq.~\eqref{eq:m_rad_sym} and for the boundary conditions mentioned above, we obtain $Q=0$ for even values of $\nu$, and $Q=1$ for odd values of $\nu$. This means that the creation or annihilation of a skyrmion with $|\nu|= 1$ can only 
occur with the breaking of the conservation of the topological charge. 
Conversely, a skyrmionium ($|\nu|=2$) can spontaneously annihilate  because~$Q=0$ for both the skyrmionium and the no-skyrmion states. We will return to this discussion in Sec.~\ref{sec:motion}.

Equation~\eqref{eq:euler_lagrange} with the specified boundary conditions is a strongly nonlinear differential equation that does not admit an explicit analytical solution and therefore should be solved numerically, for example, by the shooting method. One should solve it for different positions~$\bm{a}$ of the skyrmion center, compare the corresponding energies, and identify the local and global minima, which correspond to the metastable and stable skyrmion states at the respective positions~$\bm{a}$.

\subsubsection{Evolution equation}

The evolution of magnetization can be described by the Landau-Lifshitz-Gilbert (LLG) equation~\cite{LandauLifshitz1935,Gilbert2004},
\begin{equation}
	\partial_t\bm{m} = - \gamma \bm{m} \times \bm{H}_{\rm eff} + \alpha \bm{m} \times \partial_t\bm{m}, \label{eq:LLG}
\end{equation} 
where $\bm{H}_{\rm eff} = - M_s^{-1} \delta \mathcal{F}/\delta \bm{m}$ is the effective magnetic field, $M_s$ is the saturation magnetization, and $\gamma$ and $\alpha$ are phenomenological constants governing the precession and relaxation of the magnetization, respectively.

Numerical solution of the LLG equation on a discretized lattice, known as micromagnetic simulation (MMS)~\cite{Donahue2007}, is widely used to analyze theoretical and experimental results. MMS enables the identification of stable states and the study of time-dependent processes, such as the motion, creation, and annihilation of skyrmions. Although the LLG equation describes continuous processes and conserves the total topological charge of the magnetization in a ferromagnetic film, the discretized lattice in MMS, as well as real spin systems, breaks this continuity, allowing the creation and annihilation of skyrmions to be captured. Simulations presented in this paper were performed using the Object-Oriented MicroMagnetic Framework (OOMMF) software package~\cite{OOMMF} and the Ubermag environment~\cite{Ubermag}. Additional details are provided in Appendix~\ref{app:MF}.

Beyond MMS, which requires substantial computational resources, an analytical approach based on the LLG equation, known as the Thiele equation~\cite{Thiele1973}, can be used to study the motion of skyrmions or other localized magnetic textures. By describing the magnetization of a skyrmion using a specific shape function with parameters, such as the skyrmion center, radius, and wall thickness, one can derive an evolution equation for these parameters.

Assuming the skyrmion center is located at the point~$\bm{a}$ in a quasistable state described by Eqs.~\eqref{eq:m_rad_sym} and~\eqref{eq:euler_lagrange}, one can substitute its magnetization into the LLG equation~\eqref{eq:LLG} and derive the Thiele equation:
  \begin{equation}
    \hat{G} \dot{\bm{a}} - \alpha \hat{D} \dot{\bm{a}} + \bm{F}=0, 
    \qquad 
    \bm{F}= -\dfrac{\gamma}{4\pi M_s} \dfrac{\partial \mathcal{F}}{\partial \bm{a}}.
    \label{eq:thiele1}
  \end{equation}
  Here, $\bm{F}$ is the force induced by the electric field, while $\hat{G}$ and $\hat{D}$ are the antisymmetric gyrotropic and symmetric drag tensors, respectively, defined as
  \begin{equation}
    G_{ij} = \!\int\! \dfrac{d\bm{r}}{4\pi} \bm{m} \cdot \!\left( \dfrac{\partial \bm{m}}{\partial a_i} {\times} \dfrac{\partial \bm{m}}{\partial a_j} \right)\!,
    \;\;
    D_{ij} = \!\int\! \dfrac{d\bm{r}}{4\pi} \dfrac{\partial \bm{m}}{\partial a_i} \cdot \dfrac{\partial \bm{m}}{\partial a_j}.
    \label{eq:thiele_GD}
  \end{equation}
  This form of the Thiele equation is relatively simple, but may not accurately describe skyrmion motion in all cases. The generalized Thiele equation, which accounts for the joint evolution of such skyrmion parameters as the center, radius, domain wall thickness, and helicity, as well as explicit expressions for the tensors $\hat{G}$ and $\hat{D}$, is discussed in Appendix~\ref{app:Thiele}.

Equation~\eqref{eq:thiele1} describes the motion of a skyrmion driven by an electric field. Note that the right-hand side of Eq.~\eqref{eq:thiele1} is proportional to the gradient of the free energy, which aligns with the field gradient. Since the dissipation parameter is typically small, $\alpha \ll 1$, the velocity direction is primarily determined by the antisymmetric tensor $\hat{G}$ and the skyrmion moves nearly perpendicular to the field gradient. This phenomenon can be either an obstacle or an advantage, and we return to this discussion in Section~\ref{sec:motion}.

\subsubsection{Electric field distribution}

To proceed, we determine the specific distribution of the external electric field used to derive analytical results and perform micromagnetic simulations. Since an experimentally consistent field source is generated by a thin, electrically charged tip~\cite{Pyatakov2015}, we model it as a point charge~$q$ positioned at height~$h$ above the point $\bm{r}=-\bm{a}$ on the ferromagnetic film,  see discussion about shifting the coordinate origin to the skyrmion center near Eq.~\eqref{eq:euler_lagrange} and Fig.~\ref{fig:scheme},
\begin{equation}
	E(\bm{r})=E_0\mathcal{E}(|\bm{r}+\bm{a}|),
	\quad
	\mathcal{E}(r)=\dfrac{h\ell_w^2}{(h^2 + r^2)^{3/2}},
	\label{eq:Ez-point}
\end{equation}
where $E_0 =-2q/(\varepsilon_{f}\ell_w^2)$ is a spatially constant magnitude, and $\varepsilon_{f}$ is the dielectric permittivity of the film.
Although the realistic electric field distribution may be more complex, we expect the qualitative findings of this study to remain valid.

Hereafter, we assume that the height~$h$ is on the order of the magnetic length~$\ell_w$. Consequently, it is convenient to introduce a dimensionless effective field intensity~$\beta$ of the electric field,
\begin{equation}
	\beta=\dfrac{\gamma_{\rm ME}E_0}{2\sqrt{AK}}=\dfrac{-q\gamma_{\rm ME}}{\varepsilon_{f}\ell_w^2\sqrt{AK}}, 
\end{equation}
which can reach values on the order of $10$ in realistic experimental setups, see Table~\ref{tab:dimensional_values}. The effective DMI function $\epsilon_{\rm eff}(r)$ induced by the electric field, see Eq.~\eqref{eq:euler_lagrange}, can then be expressed as
\begin{equation}
	\epsilon_{\rm eff}(r)=\epsilon_0 + \beta\bar{\mathcal{E}}_{a}(r),
	\quad
	\bar{\mathcal{E}}_{a}(r)=\cfrac{2h\ell_w^2{\rm E}\big[4ar/\bar h^2(r+a)\big]}{\pi \bar h^2(r-a)\bar h(r+a)},
    		\label{eq:eps_a}
\end{equation}
where $a=|\bm{a}|$ represents the distance between the skyrmion center and the tip, ${\rm E}[z]$ denotes the elliptic integral of the second kind, and $\bar{h}(r)=\sqrt{h^2 + r^2}$.

\section{Stable skyrmion states\label{sec:states}}

We have performed analysis of the stable configurations of the skyrmion near the charged tip in several stages.

Firstly, we have solved the EL equation~\eqref{eq:euler_lagrange} for different distances~$a$ and skyrmion numbers~$\nu$, and identified that only coaxial configurations are stable for the chosen model of the electric field, Eq.~\eqref{eq:Ez-point}. Note that, in principle, a skyrmion can be stable in an eccentric configuration with some other form of external field even when it is radially symmetric, e.g. see Ref.~\cite{Apostoloff2024},
in which eccentric configurations of skyrmions and superconducting vortex have been predicted. 

Secondly, we systematize the coaxial configuration and identify the lowest-energy states. In particular, we reveal that in some parameter ranges the skyrmions or skyrmioniums appear to be more energetically favorable than the no-skyrmion state, which implies the possibility of skyrmion creation. 

Finally, we verify our analytical findings by means of MMS and present the phase diagram, which maps the regions of possible skyrmion states in the $a$-$\beta$ plane. 

\subsection{Skyrmions with $|\nu|=1$ at different distances $a$\label{sec:quasistable}}

We have solved the EL equation~\eqref{eq:euler_lagrange} for different distances~$a$ and skyrmion numbers~$\nu$ for certain values of effective field intensity~$\beta$ and DMI parameter $\epsilon_0$. Interestingly, each set of these parameters can produce several solutions. For comparison, there is a single solution related to a free skyrmion, i.e. for $\beta=0$, and this solution corresponds to $\nu = {\rm sgn\,}\epsilon_0$. When the absolute value of effective field intensity~$|\beta|$ is finite the multiple solutions for each $\nu=0,\pm1,\pm2,\ldots$ can appear, and the number of them generally increases with~$|\beta|$, see Appendix~\ref{sec:multi}. 
In this subsection
we focus only on skyrmions with $|\nu|=1$. The skyrmioniums ($|\nu|=2$) and target skyrmions ($|\nu|>2$) are discussed in the next subsection.

We identify quasistable skyrmion states as local minima of the total free energy $\mathcal{F}$ 
at a certain
distance~$a$ from the tip.
In addition to the energy,
it is convenient to characterize each skyrmion by its radius~$R$, defined as the distance from the skyrmion center to the outer skyrmionic wall, determined by the condition $|\theta(r = R)| = \pi/2$.

\begin{figure}[h]
	\centering
	\includegraphics[width=\linewidth]{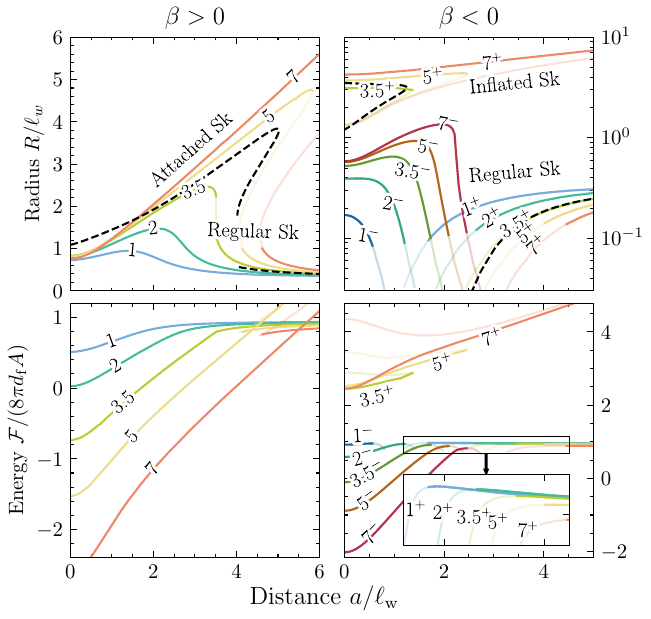}
	\caption{
        Dependences of the skyrmion radius $R(a)$ (upper panels) and free energy $\mathcal{F}(a)$ (lower panels) on the distance~$a$ from the charged tip, shown for $\nu = \pm 1$ and $|\beta| = 1,\,2,\,3.5,\,5,\,7$ (values indicated as numbers next to the curves).
    	Left panels correspond to $\beta > 0$, where only skyrmions with positive chirality ($\nu = +1$) are stable due to the positive DMI interaction.
    	Right panels correspond to $\beta < 0$; here superscripts $+$ and $-$ label the skyrmion chirality ($\nu = \pm 1$).
    	The black dashed curves show analytical approximations for $\beta = 5$ resulting from Eqs.~\eqref{eq:approximation_large} and~\eqref{eq:homogeneous_epsilon} for relevant ranges. Parameters: $\epsilon_0=0.3$, $h/\ell_w=1$.
	}
	\label{fig:FR(a)}
\end{figure}

Figure~\ref{fig:FR(a)} presents the skyrmion radius $R(a)$ (upper panels) and free energy $\mathcal{F}(a)$ (low panels) for the quasistable states with ${|\beta| = 1},\,2,\,3.5,\,5,\,7$.
The left panels correspond to $\beta > 0$, where only skyrmions with positive chirality ($\nu = +1$) exist. 
The right column shows the case $\beta < 0$, for which both chiralities $\nu = \pm 1$ are realized; these are distinguished by superscripts on the curve labels. Note that parts of the curves plotted by lighter colors corresponds to unstable skyrmion states, that can be non-minimum (saddle-point) solutions of the EL equation or dynamically instabile.

In Fig.~\ref{fig:FR(a)} one can identify different types of $R(a)$-dependence with one or several branches. A branch that we marked as ``Regular Sk'' corresponds to skyrmions with smaller radius~$R$ that are realized at relatively small field intensities, $|\beta| \lesssim 3$, or large distances to the tip, $a\gtrsim 2$. Both cases imply the weak influence of the field to the skyrmion, therefore it behaves regularly. 

Oppositely, at $|\beta| \gtrsim 3$, there are a pair of branches (stable and unstable) which grow almost linearly in radius with an increase of distance from the tip, $R\simeq a\gtrsim 2$. Skyrmions from these branches differ from regular ones in the physical nature. 
For $\beta>0$ the tip attracts the skyrmion wall making distant skyrmion increase its radius until it becomes attached to the tip.
For $\beta<0$ the tip repels the skyrmion wall making closer skyrmion inflate until it covers the tip.
Therefore, we call such skyrmions as ``Attached Sk'' for $\beta>0$ and ``Inflated Sk'' for $\beta<0$, respectively.

Note that we have found asymptotic expressions for $\mathcal{F}(a)$ and $R(a)$ in two limiting cases, (i)~${R\gg\ell_w}$ and $|\beta|\gg1$, and (ii)~${R\ll a}$, both shown in Fig.~\ref{fig:FR(a)} as black dashed lines. The first case corresponds to the ``attached'' or ``inflated'' skyrmions, cf. Fig.~\ref{fig:FR(a)}, and is based on the assumption that the skyrmion of a large radius can be considered as a cylindrical magnetic domain with radius~$R$ and a wall of thickness $\ell_w$. Therefore, its energy is approximately equal to 
\begin{equation}
	\dfrac{\mathcal{F}}{8\pi A d_F}\approx \dfrac{R}{\ell_w}+\dfrac{\ell_w}{2R}-\dfrac{\pi R}{2\ell_w}[\epsilon_0+\beta\bar{\mathcal{E}}_{a}(R)f(R-a)]. 
    \label{eq:F_R>>1}
\end{equation}
The explicit expression for the factor $f(r)$, which is the rational function of $r$, are given in the Appendix~\ref{app:asymp}. The other terms were calculated in Ref.~\cite{Wang2018}. Then the free energy achieves minimum at radius $R$ defined by the algebraic condition 
\begin{equation}
	\epsilon_0+\beta\partial_R[R\bar{\mathcal{E}}_{a}(R)f(R-a)]=[2-(\ell_w/R)^{2}]/\pi.
	\label{eq:approximation_large}
\end{equation}
Notably Eq.~\eqref{eq:approximation_large} for some values of $a$ and $\beta$ can have several solutions due to nonlinear dependence of $R\bar{\mathcal{E}}_{a}(R)f(R-a)$ on $R$. Indeed it results into existence of the branches of the inflated or attached skyrmions. 

In the second limiting case, $R\ll a$,
the skyrmion is located relatively far away from the tip. Thus the effective DMI is approximately homogeneous over whole area with nonhomogeneous magnetization,
\begin{equation}
	\epsilon_{\rm eff}\simeq\epsilon_0+ \dfrac{\beta h\ell_w^2}{(h^2 + a^2)^{3/2}}.
	\label{eq:homogeneous_epsilon}
\end{equation} 
Therefore, we can use the results obtained for the free skyrmion with the constant DMI parameter $\epsilon_{\rm eff}$, see Ref.~\onlinecite{Wang2018}.  
Equation~\eqref{eq:homogeneous_epsilon} yields approximation for branch with regular skyrmions, see. Fig.~\ref{fig:FR(a)}. As known the skyrmion is stabilized by the DMI, therefore the zero effective DMI, ${\epsilon_{\rm eff} \simeq 0}$, estimates the critical distance~$a_{\rm cr}$,
\begin{equation}
a_{\rm cr}\simeq\sqrt{|\beta h \ell_w^2/\epsilon_0|^{2/3}-h^2},   
\label{eq:a_cr}
\end{equation}
which separates branches for the regular skyrmions of positive, $\nu=+1$, and negative, $\nu=-1$, chiralities for ${\beta<0}$. Notably, that $a_{\rm cr}$ vanishes for sufficiently small ${0>\beta\gtrsim- (h/\ell_w)^2\epsilon_0}$, i.e. there is no stable skyrmions with $\nu=-1$.

The corresponding curves for both limiting cases are plotted by the black dashed lines in Fig.~\ref{fig:FR(a)} for $\beta=5$. One can see the good quantitative agreement with solutions of the EL equation plotted by the solid curves.

\subsection{Energy of coaxial configurations}

\begin{figure}[h]
	\centering
	\includegraphics[width=1\linewidth]{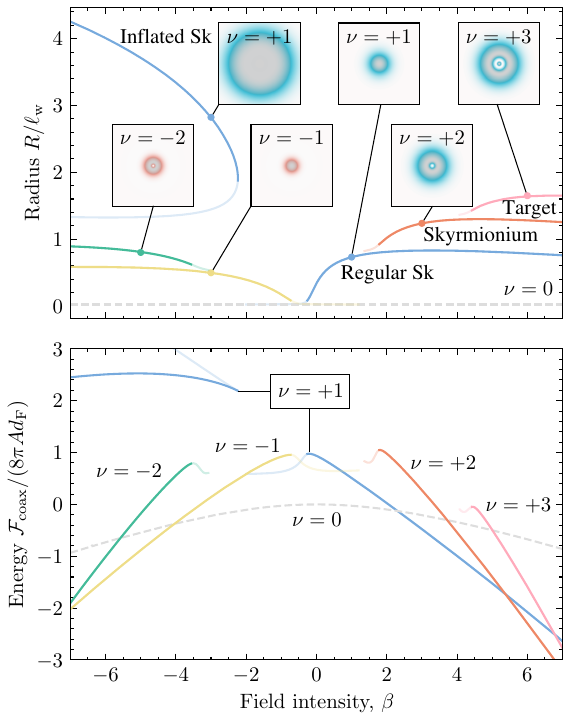}
	\caption{Dependence $R(\beta)$ and $\mathcal{F}(\beta)$ for the coaxial configurations. The upper panel shows the skyrmion radius $R$ (in units of $\ell_w$), while the lower panel presents the corresponding total free energy $\mathcal F$ normalized by $8\pi A d_F$. Solid curves indicate skyrmionic states: regular skyrmions with positive ($\nu=+1$, blue curve) and negative ($\nu=-1$, yellow curve) chirality, skyrmioniums with positive ($\nu=+2$, orange curve) and negative ($\nu=-2$, green curve) chirality, and target skyrmions ($\nu=+3$, pink curve). The lighter-colored curves denote unstable branches. The dashed gray curve corresponds to the no-skyrmion state ($\nu=0$). Insets shows the distribution of $m_z$ (in color gradient) obtained by micromagnetic simulations for each type of 
   the magnetic configurations for some value of $\beta$ marked by the point on each curve. All the insets has the same spatial scale to demonstrate the relative sizes of the
    magnetic configurations. 
        Parameters used are $\epsilon_0=0.3$ and $h/\ell_w=1$.
		}
		\label{fig:FR_coax}
	\end{figure}

From the studies of dependence $\mathcal{F}(a)$ we can conclude that the electric field in the form of Eq.~\eqref{eq:Ez-point} provides only coaxial stable configurations or repels the skyrmion from the charged tip. To systematize the possible coaxial configurations we have disregarded the high-energy states for each value of $\beta$ and identified only several stable coaxial configurations with the lowest energy. Figure~\ref{fig:FR_coax} shows the total free energy~$\mathcal{F}$ and skyrmion radius~$R$ as a function of~$\beta$ of the lowest-energy states, while insets show the characteristic magnetization profiles of 
several generic classes of coaxial solutions with $\nu=0,\pm 1,\pm 2,+3$ 
observed in our micromagnetic simulations for experimentally relevant values of $|\beta|<7$.

The blue, orange, and pink curves correspond to skyrmions ($\nu = +1$), skyrmioniums ($\nu = +2$), and target skyrmions ($\nu = +3$) with positive chirality, while the yellow and green curves represent skyrmions ($\nu = -1$) and skyrmioniums ($\nu = +2$) with negative chirality, respectively. 
The higher-order target-skyrmion configurations ($\nu \leq -3$ and $\nu \geq +4$) appear at larger values of $\beta$ than those shown in Fig.~\ref{fig:FR_coax}.
The curves plotted in the lighter colors represent the corresponding unstable (saddle-point) branches, cf. Fig.~\ref{fig:FR(a)}. 
The dashed curve denotes configurations without skyrmions ($\nu=0$) and serves as a reference, showing that there are ranges of~$\beta$ 
where the skyrmion or skyrmionium 
becomes energetically favorable.

In particular, for DMI parameter $\epsilon_0=0.3$ and height~$h=\ell_w$ used in Fig.~\ref{fig:FR_coax}, the skyrmions of positive (or negative) chirality has lower energy than non-skyrmion states when $\beta>\beta_+\approx 2.2$ (or $\beta<\beta_-\approx -4.1$), while skyrmioniums of positive chirality favors $\beta>\beta_{+2}\approx 5.4$.
Skyrmioniums of negative chirality and target skyrmions may be energetically efficient for the higher intensities than are experimentally relevant, $\beta>7$. 
It shows that skyrmion or skyrmionium, in principal, can be created or annihilated by the electric field of certain magnitude. 

\subsection{Phase diagram~\label{sec:diagram}}

\begin{figure}[h]
	\centering 
	\includegraphics[width=0.9\linewidth]{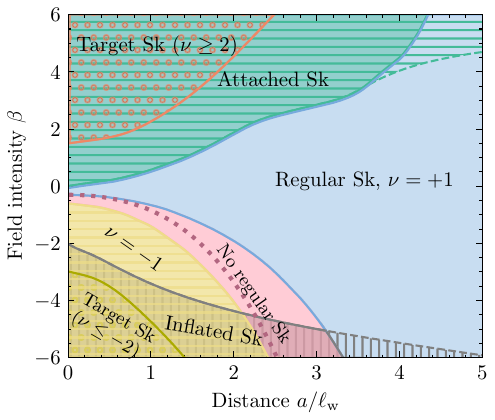}
	\caption{
		Phase diagram for the quasistable skyrmion states in the plane of distance~$a$ to the tip and field intensity~$\beta$, obtained by solving the EL equation~\eqref{eq:euler_lagrange}. For positive chirality, $\nu = +1$, three regions can be distinguished: regular, attached, and inflated skyrmions. 
        The regions of attached and inflated states bounded by the solid and dashed curves on either side 
        correspond to states with greater energy and are inaccessible in MMS. 
        Orange and green region correspond to target skyrmions of positive and negative chirality, respectively.
        The purple dotted curve is defined by condition $\epsilon_{\rm eff}=0$ in Eq.~\eqref{eq:homogeneous_epsilon} and separates the areas of skyrmions with negative, $\nu = -1$, and positive, $\nu = +1$, chirality. 
        The solid curves bound areas of parameters, for which attached or inflated states can be observed in MMS. Parameters: $\epsilon_0 = 0.3$, $h/\ell_w = 1$.
    }	
        \label{fig:diagram}
\end{figure}

To verify results obtained analytically we perform a series of micromagnetic simulations. These simulations confirm the possibility to stabilize skyrmion or skyrmionium for the correspondent ranges of $\beta$.
Summarizing the results obtained analytically and verified by numerical micromagnetic simulations we plot Fig.~\ref{fig:diagram}, the diagram presenting the regions of different possible skyrmionic configurations in parameter space of the distance $a$ and the electric field intensity $\beta$.

The blue area corresponds to regular skyrmion states of positive chirality, $\nu=+1$, that influenced by the electric field weakly, whether $\beta$ is small or distance $a$ is great. The MMS confirms that they retain their radial symmetric shape and relatively slow moving to or from the tip, whereas 
circular orbiting motion dominating over radial attraction or repulsion, see Section~\ref{subsec:moving_skyrmion} for details.

Configurations with inflated ($\beta<0$) and attached ($\beta>0$) skyrmions of positive chirality, $\nu=+1$, occupy the gray vertical-hatched and green horizontal-hatched areas, respectively. Note that these areas has two types of boundaries: solid lines indicate the boundary of the regular-skyrmion area and the end of the regular-skyrmion branches from Fig.~\ref{fig:FR(a)}, while the dashed curve corresponds to the end of the inflated- or attached-skyrmion branches there. The MMS shows that the inflated and attached states outside the solid border lines are not possible because the regular skyrmions are stabilized there. Additionally should be noted that inflated (unlike attached) skyrmions are stable only if their domain wall is wrapped around the tip, see details in subsection~\ref{subsec:resize_skyrmion}. 
	
	An area of skyrmions of negative chirality ($\nu=-1$) is located only for $\beta < 0$ and marked by yellow shadowing. The pink region denotes the absence of stable skyrmion states. The purple dotted curve represents approximate boundary between skyrmions of positive and negative chirality described by Eq.~\eqref{eq:a_cr}. Physically, this condition means that the effective DMI vanishes, $\epsilon_{\rm eff}=0$, cf. Eq.~\eqref{eq:homogeneous_epsilon}, and skyrmions cannot be stabilized.

    The orange and green dotted areas show the possible target skyrmion states with $\nu\geq+2$ and $\nu\leq -2$, respectively. 
    Their boundaries are given approximately 
    basing on the assumption that the target skyrmions are stable only when the outer skyrmion wall covers the tip, which is consistent with the MMS.

\section{Moving, resizing, creating, and annihilating skyrmions~\label{sec:motion}}

In this section, we discuss the manipulation of skyrmions 
using a non-uniform electric field 
created by a charged tip. 
Based on the results of the previous Section, 
we can conclude that this field can
\begin{itemize}  
\item[(i)] drive skyrmions to move, because the free energy depends on the distance between the skyrmion and the tip, see the lower panels of Fig.~\ref{fig:FR(a)};
\item[(ii)] create/annihilate skyrmion or skyrmionium switching to appropriate ranges of field intensity $\beta$, where the energy of skyrmion or skyrmionium appears to be lower/higher than some other configurations, see the lower panel of Fig.~\ref{fig:FR_coax};
\item[(iii)] resize skyrmions located under the tip, because their radius depends on the field intensity, see the upper panel of Fig.~\ref{fig:FR_coax}.
\end{itemize}

The possibility to create and annihilate skyrmion or skyrmionium is 
based on the results of MMS. Indeed, due to the discrete nature of the real spin system as well as the model of macrospins used in MMS, see details in Appendix~\ref{app:MF}, the topological charge~$Q$, see Eq.~\eqref{eq:rho_Q}, may not be conserved in the creation or annihilation process.

\subsection{Moving skyrmions\label{subsec:moving_skyrmion}}

Below we will discuss how to drive the skyrmions by the electric field. We focus on the skyrmion of the positive chirality, $\nu=+1$, because only such skyrmions can be stable without electric field for $D>0$. The motion of the skyrmion can be described numerically by the MMS and analytically by the Thiele equation, which we employ in two formulations.

The first form, see Eq.~\eqref{eq:thiele1}, describes the motion of the skyrmion center, while its shape is assumed to be taken from Eq.~\eqref{eq:euler_lagrange} in each time moment for the corresponding value of distance~$a$. The second form is more general
and governs the evolution of the skyrmion center, radius, wall thickness, and helicity, see  Apendix~\eqref{app:Thiele}. Note that the second form are preferably for the precise description of the complicated motion because the evolution of radius are related to the evolution of helicity, see Eq.~\eqref{eq:Thiele2}. However, the first form gives the qualitative description of the motion basing on the quasistable states discussed in subsection~\ref{sec:quasistable} and shown in Fig.~\ref{fig:FR(a)}. Below we provide such a description.

Due to the force acting from a single charged tip, a skyrmion generally moves along the spiral trajectory, attracting (clockwise) or repelling (counterclockwise), nearly perpendicular to the field (and energy) gradient, because its velocity direction is 
mainly determined by the antisymmetric tensor $\hat{G}$. 
Interestingly, the velocity varies significantly depending on the distance from the tip. 
In particular, the energy gradient, see Fig.~\ref{fig:FR(a)}, has a different scale: skyrmions that are closer than a critical distance  to the tip, $a<a_{\rm cr}\sim 4\ell_w$, move several times faster than those at 
larger distances, $a>a_{\rm cr}$. Therefore, it may be reasonable to use multiple tips to control a skyrmion
as discussed in Section~\ref{sec:platform}.

\begin{figure}[h]
	\centering
	\includegraphics[width=1\linewidth]{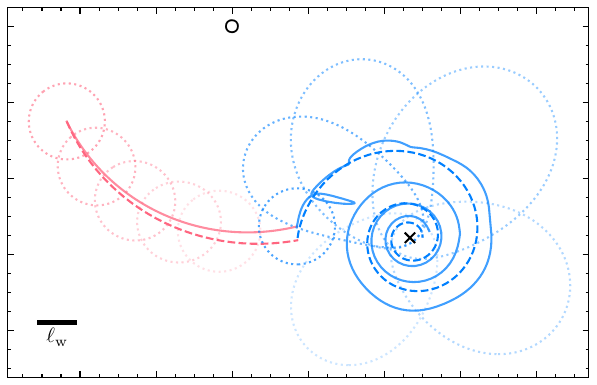}
    \caption{
		Trajectories of 
        the topological center of skyrmion, 
        $\bm{r}_{\rm tc}=\int d^2\bm{r}
    \,\bm{r}\rho(\bm{r})$, cf.~Eq.~\eqref{eq:rho_Q},
    driving by a distant tip denoted with ``$\circ$'' (red arcs) and by a closer tip denoted with ``$\times$'' (blue spirals) and  calculated by means of Thiele equation
        Eq.~\eqref{eq:thiele1} (dashed curves), and extracted from the MMS (solid curves). 
		The lighter dotted curves are snapshots of the skyrmion walls for some intermediate time moments. 
		Parameters: $\epsilon_0=0.5$ and $\alpha = 0.05$ for both curves, and $\beta = \beta_\circ=-3$ for the distant tip, $\beta = \beta_\times=2$ for the closer tip.
    }
	\label{fig:traject}
\end{figure}

In Fig.~\ref{fig:traject} we provide the comparison of the skyrmion trajectories obtained from MMS (solid curves) and using Thiele equation (dashed curves). We consider two basic types of trajectories, arc-like motion (red curves) around the distant charged tip (``$\circ$'' mark) and spiral motion (blue curves) around the nearby tip (``$\times$'' mark). The idea behind Fig.~\ref{fig:traject} is as follows. We switch on the voltage at the ``$\circ$''-tip and pull the skyrmion closer to the switched-off ``$\times$''-tip.
Then the voltage on the ``$\times$''-tip is applied and the skyrmion starts to be attracted to the ``$\times$''-tip, while ``$\circ$''-tip is turned off. 

We can see that 
for physically reasonable conditions, $\alpha = 0.05$, $\epsilon_0=0.5$, $\beta_\circ=-3$ and $\beta_\times = 2$, used in Fig.~\ref{fig:traject}, cf. Table~\ref{tab:dimensional_values}, the Thiele equation give good qualitative and satisfactory quantitative agreement with the results of MMS even in spite of the deviation of the shape of the skyrmion from being centrally symmetric.
Note that the both forms of Thiele equation described above give approximately the same trajectories, so we present only one resulted from Eq.~\eqref{eq:thiele1}. This
justifies the use of the Thiele equations for estimating the skyrmion position and the characteristic times in different regimes of motion which can be used in the different scenarios for skyrmion manipulation discussed in Sec.~\ref{sec:platform}. 

\subsection{Creation of magnetic objects}

\subsubsection{Skyrmions of positive chirality}

Here we introduce two methods for the creation of a skyrmion of positive chirality ($\nu=+1$) mediated by the electric field of the tip. 

The first approach relies on a rapid reversal (``swing'') of the field polarity. Initially, in the homogeneously magnetized ferromagnet film, $\bm{m}=\bm{e}_z$, a strong enough electric field with $\beta = -\beta_{+\rm sw}<0$ 
induces a localized perturbation, whose magnetization 
resembles a coaxial skyrmion of positive chirality, $\nu=+1$, except the region close to the tip, where again $\bm{m}\approx\bm{e}_z$.
When the field is rapidly switched to the opposite polarity, $\beta = \beta_{+\rm sw}>0$, the region near the tip, where $\bm{m}\approx\bm{e}_z$, shrinks to the size of several spins (or macrospins used in MMS). This state is energetically unfavorable, magnetization near the tip turns over, $\bm{m}\approx-\bm{e}_z$, breaking the conservation of topological charge, and a skyrmion with $\nu=+1$ is created. 

	In alternative (``direct'') approach, skyrmion creation can occur without the field swing. One can use 
    an abrupt switching of electric field from the uncharged state, $\beta = 0$, to sufficiently strong $\beta>0$. 
    A magnetization configuration initially forming at $\beta >0$ resembles a coaxial skyrmion of negative chirality, $\nu=-1$, except for the region close to the tip. For sufficiently large $\beta = \beta_{+\rm dir}>0$ this 
    region shrinks to the size of several spins, which is energetically unfavorable.
    Therefore, the skyrmion 
    can be created, breaking the conservation of topological charge. However, the skyrmion with $\nu=-1$ cannot be stable at $\beta>0$ and, therefore, the skyrmion with $\nu=-1$ has to transform into a skyrmion with $\nu=+1$ via the rotation of its magnetization (changing helicity), now conserving the topological charge. 
The definition of the helicity is given in Appendix~\ref{app:Thiele}.
    
    We note that in the ``direct'' approach a higher field  intensity is required than in the ``swing'' method, ${\beta_{+\rm dir}>\beta_{+\rm sw}}$, because there is no preliminary perturbation of the magnetization, which helps skyrmion 
    with $\nu=+1$ 
    to be nucleated.
    We estimate the values of the required electric fields from the MMS: 
    $\beta_{+\rm cr}\simeq 7.6$ and $\beta_{+\rm sw}\simeq 4.4$, see Appendix~\ref{app:MF}.

\subsubsection{Skyrmions with negative chirality\label{subsec:create_skyrmion:nu<0}}

For the skyrmions with negative chirality, $\nu=-1$, the same two (``direct'' and ``swing'') methods can be applied with the same qualitative results as for $\nu=+1$. 
However, due to the positive DMI parameter, $\epsilon_0>0$, the values of the critical fields that are needed for the creation of a skyrmion with $\nu=-1$ appear to be slightly stronger than the field for $\nu=+1$, cf. the energies for $\nu=-1$ and $\nu=+1$ in the lower panels of Fig.~\ref{fig:FR_coax}.
Specifically, the correspondent critical values are 
${\beta_{-\rm dir} \simeq -8.0}$ and ${\beta_{-\rm sw} \simeq -5.4}$.

\subsubsection{Skyrmioniums and target skyrmions}

Similarly to a regular skyrmion, skyrmioniums and target skyrmions, $|\nu|\geq2$ can be created with application of electric fields of higher intensity. In MMS we observed creation of target skyrmions when $|\beta| \gtrsim 8$. 

Note that the appropriate field should be applied not only for creating such target skyrmions, but also to keep them, because they are not stable (deforming or/and annihilating) in the absence of the field.
Moreover, 
the slow change the high field intensity allows for precise control of the number of domain walls, e.g. $|\nu|$, in target skyrmions. 

As known, in most cases, skyrmioniums are created randomly or within a bunch across a magnetic material surface~\cite{Vizarim2025} and appear to be metastable~\cite{nakamura2024}.
Remarkably, using the electric field one can not only  
create target skyrmions in controllable way, but also stabilize them. 

\subsection{Annihilation of magnetic objects}

\subsubsection{Skyrmions of positive chirality\label{subsec:ann:nu>0}}

As one can see from Fig.~\ref{fig:diagram} the skyrmion of the positive chirality, $\nu=+1$, shifted from the tip is stable practically for any intensity of electric field. Therefore, to be annihilated, the skyrmion should be firstly attracted to the tip nearly coaxially, see Fig.~\ref{fig:traject} and details in Sec.~\ref{subsec:moving_skyrmion}. Then one should switch the voltage on the tip so that the skyrmion wall appears within the region with approximately zero effective DMI, $\epsilon_{\rm eff}(a, r \simeq
R) \simeq 0$. In the case of a nearly coaxial skyrmion, assuming radius $R$ to be large and using Eq.~\eqref{eq:eps_a}, we can estimate the required field intensity as
\begin{equation}
\beta_\mathrm{an} \simeq - \epsilon_0 \frac{(h^2 + R^2)^{3/2}}{h \ell_w^2}.
\label{eq:anihilation_beta_boundary}
\end{equation}

Note that the skyrmion of $\nu=+1$ can be repelled from the tip, cf.~subsection~\ref{subsec:moving_skyrmion}, inflate, cf.~subsection~\ref{subsec:resize_skyrmion}, or even change chirality, if the electric field is substantially weaker or stronger than $\beta_\mathrm{an}$, because the domain walls of the skyrmion are unstable only for the zero DMI. 

\subsubsection{Skyrmions of negative chirality\label{subsec:ann:nu<0}}

Unlike skyrmions of positive chirality, skyrmions of negative chirality, $\nu =-1$, 
require negative field intensity to exist.
However, turning on the positive field intensity $\beta > 0$ may lead to switch the chirality of a skyrmion from negative to positive, avoiding annihilation. 
Therefore, to annihilate a skyrmion reliably we should apply the same field $\beta_{\rm an}$ as discussed in previous subsection.

\subsubsection{Skyrmioniums and target skyrmions}

Because skyrmioniums, $|\nu|=2$, and target skyrmions, $|\nu|>2$, are not stable without additional condition, in particular, the appropriate electric field, then it is enough to turn the field off to destroy them. 

Also the tip and skyrmion or target skyrmion can be pulled apart somehow at sufficient distance, see the correspondent areas in the phase diagram, Fig.~\ref{fig:diagram},
then they annihilate spontaneously. 

\subsection{Resizing skyrmions\label{subsec:resize_skyrmion}}

Some experimental applications of skyrmion-based devices may require flexible control over the size of skyrmions. For example, the mentioned above skyrmion-vortex pairs in hybrid ferromagnet-superconductor heterostructures can be coaxial or eccentric depending on the size of the skyrmion~\cite{Andriyakhina2021}. Moreover, the spatial structure of the Majorana states being hosted in such pairs also depends on the size of skyrmion~\cite{Rex2019}. 

The electric field can be used to adjust the size of a skyrmion with positive chirality, $\nu=+1$, within a relatively wide range. Indeed, one can see from Fig.~\ref{fig:FR_coax} that the coaxial skyrmion can be shrunk nearly to $R\simeq0$ for small negative values of $\beta$ and slightly enlarged to $R\sim\ell_w$ for $\beta>0$. 

Additionally, large enough negative values of $\beta$ can cause the skyrmion to inflate to large values $R\sim 3\ell_w$, see Fig.~\ref{fig:FR_coax}.
However, upon switching the electric field to negative values, the skyrmion does not necessarily inflate immediately. Instead, for sufficiently strong or weak negative fields, the skyrmion can become unstable and be destroyed. More precisely, stability requires that the skyrmion radius 
satisfies the condition $\epsilon_{\rm eff}(a, r \simeq R) > 0$. 
For a nearly coaxial skyrmion with large enough radius $R$, this yields an estimate for the required field intensity as $\beta_{\rm in}>\beta>\beta_{\rm an}$, where $\beta_{\rm in}$ is the right limit of the ``Inflated Sk'' branch and $\beta_{\rm an}$ is given by Eq.~\eqref{eq:anihilation_beta_boundary}. 
Thus, after switching to $\beta<0$, only skyrmions with large enough radius can survive and subsequently inflate. Therefore, this regime can be realized only for sufficiently strong DMI, $\epsilon_0 \gtrsim 0.4$. Moreover, for $\epsilon_0 \gtrsim 0.47$ the situation changes qualitatively, as the branches of regular and inflated skyrmions turn out continuously connected. In this case, the transition becomes smooth and can be achieved by a gradual change of the electric field from positive to negative values without destruction of skymion.

In principle, the size of the skyrmion of negative chirality, $\nu=-1$, can be adjusted in similar ways by varying negative $\beta$, see Fig.~\ref{fig:FR_coax}. 
However, such skyrmions can not be inflated by switching to positive $\beta$.

\section{Platform for the skyrmion manipulation\label{sec:platform}}

In this section we will present the conceptual design for a platform that allows for the 
manipulation of skyrmions and describe several possible applications 
of 
the use of the proposed scheme. 

\begin{figure}[t]
	\centering
	\includegraphics[width=0.9\linewidth]{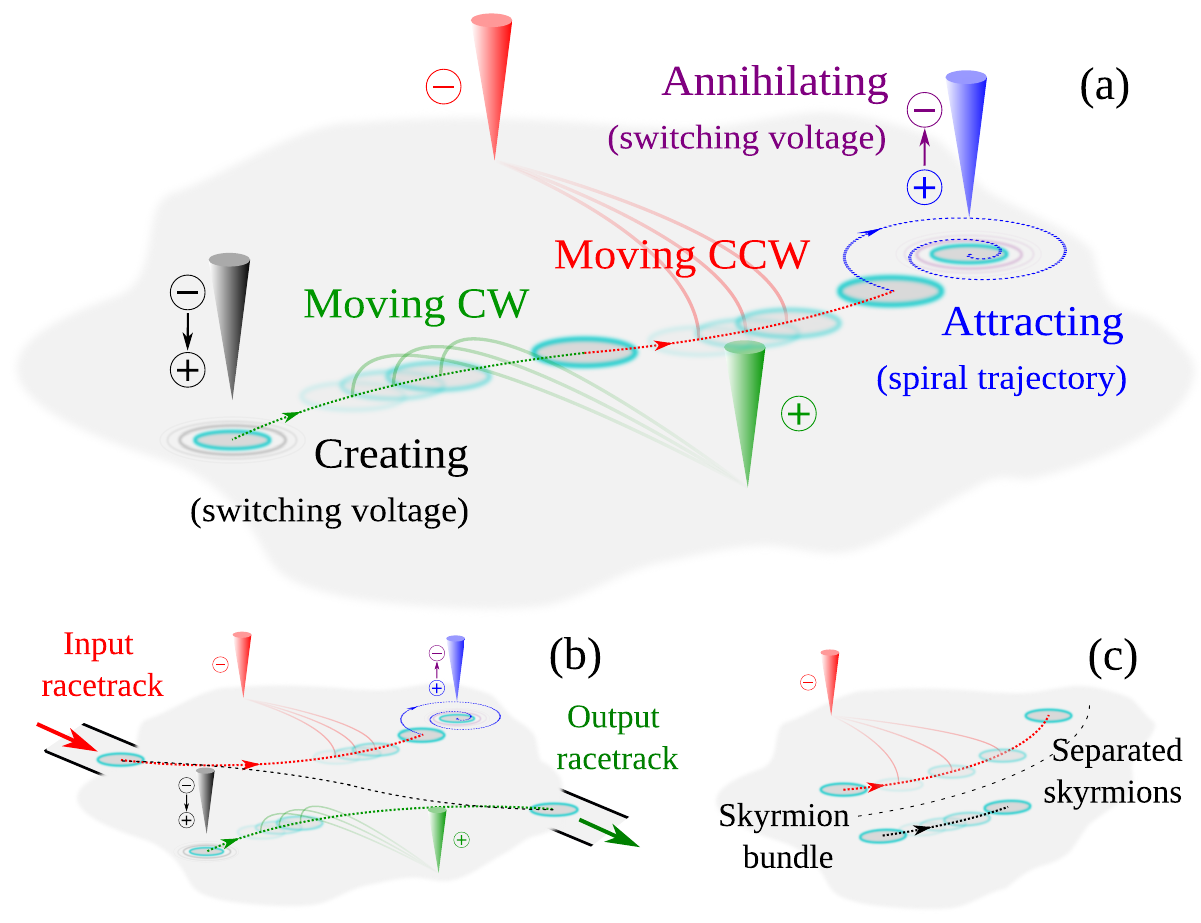}
    \caption{Platform for the skyrmion manipulation. Several tips sequentially switch voltage to create, move, and annihilate a skyrmion. Panels illustrates different scenarios: (a) internal manipulation; (b) external manipulation; (c) separation of skyrmions.
    }
	\label{fig:platform}
\end{figure}

The base of the platform consists of the ferromagnetic film and $N$ stationary tips distributed at the distance 
{$l_{\rm tips}\approx 7\ell_w$ from each other,} 
which enables the guided movement of a skyrmion between any two points within the area $S\sim N l_{\rm tips}^2$. To pull skyrmion in the required direction one should apply different voltages to the three (or less) nearest tips. The exact values of the required voltages can be found, for example, by solving Thiele equation, Eqs.~\eqref{eq:thiele1} or~\eqref{eq:Thiele2}. The minimal number of tips required is $N=4$, see Fig.~\ref{fig:platform}. Such number allows us to move a skyrmion in any desired direction, even if it is located near some tip by applying voltage to the other three tips. 
Below consider three practical scenarios of skyrmion's control that can be realizes by means of the proposed platform. 

\textit{Internal manipulation.} The skyrmion, created by a tip, can be guided along a desired trajectory by applying voltage to the appropriate tips  (see Fig.~\ref{fig:platform}a).  Once the skyrmion has been used, it can be annihilated.

\textit{External manipulation.} Another scenario can be realized if the platform is connected to several outputs/inputs of skyrmions, e.g. racetracks~\cite{Tomasello2014,He2023} or other devices (see Fig.~\ref{fig:platform}b).  When the skyrmion appears from one of the inputs it should be guided to a certain output. For that purpose the tips should be set up closer than $l_{\rm tips}/2$ to the outputs/inputs and then they can guide the skyrmion to the 
proper position as discussed in the previous scenario. Additionally, the skyrmion arriving from the input can be annihilated if it should be discarded or a skyrmion can be created and sent to the output. 

\textit{Separation.} In this scenario, we use the tips to separate a single skyrmion from the 
skyrmion bunch created by some other methods. When applied, the electric field affects all the skyrmions, but with different forces. Moreover, the skyrmions which are closer than $l_{\rm tips}/2\approx4\ell_w$ to the tip are moving with qualitatively higher velocity than more distant ones, cf. the energy gradients $\partial_a \mathcal{F}$ for the corresponding ranges in the left lower panel in Fig.~\ref{fig:FR(a)}. Therefore, we can separate 
nearby skyrmions from the distant ones (see Fig.~\ref{fig:platform}c).  Performing such a separation sequentially by using different tips, one can separate a single skyrmion from the others or even isolate all the skyrmions.

\section{Discussions and conclusions\label{sec:discuss}}

In this work, we propose an energy-efficient and precise tool for the flexible manipulation of Néel-type skyrmions in ferromagnetic films. This method is based on the magnetoelectric effect induced by a localized electric field generated by one or several charged tips. Combining complementary numerical micromagnetic simulations and analytical approaches, Euler-Lagrange and Thiele equations, we develop a consistent theory describing the stability and dynamics of skyrmions under the 
effect of the electric field. We identify several distinct dynamical regimes of skyrmion motion near the tip and map them onto a phase diagram, see Fig.~\ref{fig:diagram}.

Moreover, we predict analytically and demonstrate numerically that the electric field can create and annihilate skyrmions of both chiralities, as well as more complex magnetic textures such as skyrmioniums and target skyrmions. Our estimates show, see Table~\ref{tab:dimensional_values}, that characteristic magnitude of electric field allowing creation of a skyrmion is $E_0\sim 1$~MV/cm that can be reachable locally near thin charge tip in experiments~\cite{Pyatakov2015,Pyatakov2015-ufn,Wang2019,Wang2022}.

In addition, we discuss the principles of a platform, see Fig.~\ref{fig:platform}, that consists of a ferromagnetic film and an array of metal tips that can guide the motion, creation, and destruction of individual skyrmions by applying voltage to the tips. We also propose several scenarios for practical use of this platform, such as separating individual skyrmions from a group and manipulating them inside or outside the platform for further processing.

The developed approach obviously holds great promise for classical information processing based on skyrmion manipulation. Additionally, the skyrmions are predicted to assist the existence~\cite{Yang2016,Rex2019} of the Majorana electronic state in hybrid ferromagnet-superconductor heterostructures. This opens up opportunities  
to use the platform for realizing fault-tolerant quantum computation, as an alternative to those proposed in Refs.~\cite{Nothhelfer2022,Konakanchi2023}.  In Ref.~\cite{Rex2019} it was shown that if the skyrmion topological charge $Q$ is odd, the skyrmion should be accompanied with the coaxial superconducting vortex in order to host the Majorana states. Therefore, it will be of practical interest to extend our theory to the triple configurations: vortex-skyrmion-tip. In particular, an electric field can change the size of skyrmion and, therefore, can 
couple or disentangle the vortex and skyrmion influencing or even destroying the Majorana state, because both coaxial and displaced configurations of skyrmion-vortex can exist~\cite{Andriyakhina2021,Andriyakhina2022,Apostoloff2024}.

Moreover, our theory can be extended in the following ways. Firstly, it is interesting to study the possibility of creating and manipulating the high-order skyrmions, in particularly, with even topological charge, e.g. $|Q|=2$, is predicted~\cite{Yang2016} to provide existence of the Majorana state in superconductor without the vortex. For the high-order skyrmions, the DMI should be nonhomogeneous not only radially, but 
azimuthal~\cite{Niu2025}, that can be achieved by a pair of tips of different voltages. 

Secondly, 
it is interesting to study the motion of the Bloch-type skyrmions near a charged tip. Due to the specific orientation of the magnetization in the Bloch-type skyrmion the electric field does not affect them if use Eq.~\eqref{eq:F_ME_DMI} naively. However, the effective DMI induced in the ferromagnetic film by the magnetoelectric effect should restructuring the skyrmion magnetization, making them not exactly Bloch-type, and, then, it can be attracted to or be repelled from the tip. Moreover, the electric field can be used to change the type of the skyrmion from Bloch- to quasi-N\'eel-type and, therefore, to obtain the Majorana state in ferromagnets with Bloch-type skyrmions.

\begin{table}
	\centering
	\begin{tabular}{|c|c|c c|}
		\hline
		Value & Notation & \multicolumn{2}{|c|}{Value} \\
		\hline
		Exchange coefficient    & $A$           & $10^{-11}$    & $\text{J/m}$  \\
		Anisotropy coefficient  & $K$           & $10^{3}$      & $\text{J/m}^3$\\
		DMI             
             & D             & $10^{-4}$     & $\text{J/m}^2$\\
		Domain wall length      & $\ell_w$    & $10^{-7}$     & $\text{m}$    \\
		Field intensity ($\beta = 1$) & $E_0$    & $10^8$        & 
        $\text{V/m}$  \\
		\hline
	\end{tabular}
	\caption{Dimensional values by orders of magnitude.}
	\label{tab:dimensional_values}
\end{table}

\begin{acknowledgments}
The authors are grateful to A.~Buskina for the collaboration on the initial stage of the project. 
The authors acknowledge useful discussions with E.~Andiyakhina and A.~Pyatakov. One of us (I.S.B.) is indebted to R. Mamin for drawing his attention to the magnetoelectric effect in chiral ferromagnets. 
The work was funded in part by the Russian Science Foundation under the grant No.~24-12-00357 (quasi-stable states and dynamics of skyrmion described by the Euler-Lagrange and Thiele equations and platform for the skyrmions manipulations), by Ministry of Science and Higher Education (numeric micromagnetic simulations), and by Basic research program of HSE under Grant No. HSE-BR-2025-57 (methods for the creation and annihilation of skyrmions and skyrmioniums).
The authors acknowledge personal support from the Theoretical Physics and Mathematics Advancement Foundation ``BASIS''. 
We acknowledge the computing time provided to us at computer facilities at L.D.~Landau Institute for Theoretical Physics.
\end{acknowledgments}

\begin{center}
 \bf{DATA AVAILABILITY}  
\end{center}

The data that support the findings of this article are not
publicly available. The data are available from the authors
upon reasonable request.

\appendix

\section{Free skyrmion and domain wall ansatz\label{app:DW}}
	
    The magnetic energy of the ferromagnet film without any external forces is presented in Eq.~\eqref{eq:MagFe}. It allows the existence of the free skyrmions, which radially symmetric magnetization is described by Eq.~\eqref{eq:m_rad_sym}. Minimizing the energy we get the Euler-Lagrange equation~\eqref{eq:euler_lagrange}, where $\epsilon_{\rm eff}=\epsilon_0$ is spatially uniform DMI parameter. The exact solution can be found in the same way as described after that equation, e.g. by shooting method. However, it is shown in Ref.~\cite{Wang2018} that a good approximation for the free skyrmion angle~$\theta(r)$ is   
    \begin{equation}
      \theta_{R\delta}(r)
      =2\arctan\dfrac{\sinh(R/\delta)}{\sinh(r/\delta)},
      \label{eq:DW}
    \end{equation}
	which is known as $360^\circ$ domain wall (DW) ansatz. 
	
	Note several important features of DW ansatz. Firstly, the parameters $|R|$ and $\delta$ mean the radius and the effective domain wall thickness of the skyrmion, while $\chi={\rm sgn}{(R)}$ reflects the chirality. Secondly, these parameters can be obtained by direct minimization of the free energy, Eq.~\eqref{eq:MagFe}, with Eqs.~\eqref{eq:m_rad_sym} and~\eqref{eq:DW} substituted. The resulting function $\theta_{R\delta}(r)$ approximates the exact solution of the EL equation~\eqref{eq:euler_lagrange} within only $1-2\%$ of deviations~\cite{Wang2018}. 
	
	Thirdly, when the skyrmion radius is large, $|R|\gg1$, the expression~\eqref{eq:DW} reduces to simple nearly straight domain wall,
    \begin{equation}
      \theta_{R\delta}(r)
      \approx \chi\arccos\tanh[( r -|R|)/\delta],
      \label{eq:DW_R>>1}
    \end{equation}
which form a circle of radius $|R|$.
Moreover, the skyrmion wall $\delta$ obviously tends to the thickness~$\ell_w$ of straight domain wall, i.e. $\delta\approx\ell_w$.

    \section{Generalized Thiele equation\label{app:Thiele}}
    
    Here we present the generalized Thiele equation describing the dynamics of a skyrmion whose magnetization is parametrized by the collective coordinates ${\bm{\xi}=(a,\varphi,R,\delta,\eta)}$. They are position of the skyrmion's center~$\bm{a}$, given in polar coordinates by the radial distance~$a$ and azimutal angle~$\varphi$ relative to the origin, radius~$R$, effective domain wall thickness~$\delta$, and helicity~$\eta$. 
    Supposing the skyrmion angle can be approximated by the domain wall ansatz, i.e. $\theta(r)= 
    \theta_{R\delta}(r)$, see Eq.~\eqref{eq:DW}, we write the magnetization as follows,    
    \begin{eqnarray}
    	\bm{m}[\bm{\xi}]&=&\bm{e}_z \cos \theta_{R\delta} (r_{\bm{a}})
    	\notag\\
    	 && {}+ (\bm{e}_{r_{\bm{a}}}\cos\eta+\bm{e}_{\phi_{\bm{a}}}\sin\eta)\sin \theta_{R\delta} (r_{\bm{a}}),
         \label{eq:m_radial_helicity}
    \end{eqnarray}
	where $\bm{e}_{r_{\bm{a}}}$ and $\bm{e}_{\phi_{\bm{a}}}$ means the unit vectors in the radial and azimuthal directions of the cylindrical coordinate system with the origin $\bm{a}$, and $r_{\bm{a}}=|\bm{r}-\bm{a}|$.  
	
	Substituting the skyrmion magnetization~\eqref{eq:m_radial_helicity} into LLG equation~\eqref{eq:LLG}, multiplying it on $\partial \bm{m}/\partial{\xi_i}$, and integrating over whole ferromagnetic film, we derive the generalized Thiele equation in the same form as presented into Eqs.~\eqref{eq:thiele1} and~\eqref{eq:thiele_GD}, but $\bm{a}$ should be changed to $\bm{\xi}$.

	Below we calculate the explicit expression for the non-zero components of tensors $\hat{D}$ and $\hat{G}$,
	\begin{equation}
		\label{eq:GDviaQ}
	\begin{aligned}
		&G_{a,\varphi}=-G_{\varphi,a}=a,
        \quad
		\\
		&G_{R,\eta}=-G_{\eta, R}=q_0q_1\ell_w,
		\quad
		G_{\delta,\eta}=-G_{\eta, \delta}=q_3\ell_w,
        \\
		&D_{a,a}=D_{\varphi,\varphi}/a^2=q_2, 
		\quad
		D_{\eta,\eta}=q_1\ell_w^2,
		\\
		&D_{R,R}=q_0^2q_1,
		\quad
		D_{R,\delta}=D_{R,\delta}=q_0q_3,
		\quad
		D_{\delta,\delta}=q_4.    
	\end{aligned}
	\end{equation}
	Here we 
    introduce functions $q_i(R,\delta)$:
	\begin{subequations}
    \label{eq:qi}
	\begin{eqnarray}
		q_0&=&\dfrac{\ell_w}{\delta}\coth\dfrac{R}{\delta},
		\quad
		q_{1}=\dfrac{1}{2}\int_0^{\infty} \dfrac{dr\, r}{\ell_w^2} \sin^2\theta_{R\delta},
		\\
		q_{2}&=&\dfrac{1}{4}\int_0^{\infty} dr\, r\Big[
		(\partial_r\theta_{R\delta})^2+\dfrac{\sin^2\theta_{R\delta}}{r^2}
		\Big],
		\\
		q_{3}&=&-q_0q_1\frac{ R}{\delta }+\dfrac{2\ell_w}{\delta}\int_0^{\infty} \dfrac{dr\, r}{\ell_w^2} \sin^2\dfrac{\theta_{R\delta}}{2},
		\\
		q_{4}&=&-q_1\frac{ (Rq_0)^2+\ell_w^2}{\delta ^2}-2 q_0q_3\frac{ R}{\delta }	
		\\
		&&{}+\dfrac{\ell_w^2}{2\delta^2}\int_0^{\infty} \dfrac{dr\, r^3}{\ell_w^2}\Big[
		(\partial_r\theta_{R\delta})^2+\dfrac{\sin^2\theta_{R\delta}}{r^2}
		\Big].
		\qquad
	\end{eqnarray}
	\end{subequations}
	
	The free energy $\mathcal{F}=\mathcal{F}_{\text{magn}}+\mathcal{F}_{\text{ME}}$, see Eqs.~\eqref{eq:MagFe}  and~\eqref{eq:F_ME_DMI}, can also be expressed via the functions $q_i(R,\delta)$,
	\begin{equation}
    \label{eq:F-via-q}
		\mathfrak{F}=\dfrac{\mathcal{F}}{8\pi Ad_F}=
		(q_1+2q_2+p\cos\eta)/2,
	\end{equation}
where $p(a,R,\delta)$ is
\begin{eqnarray}
	p&=&\int_0^{\infty} \dfrac{dr\, r}{\ell_w}\Big
	(\partial_r\theta_{R\delta}+\dfrac{\sin2\theta_{R\delta}}{2r}
	\Big)[\epsilon_0-\beta\bar{\mathcal{E}}_{a}(r)].
	\qquad
\end{eqnarray}
Then the force $\bm{F}\propto\partial\mathfrak{F}/\partial\bm{\xi}$ can be calculated straightforward.

Its instructive to introduce a new collective coordinate~$\rho(R,\delta)$ that can be called ``adjusted radius'', see also Eq.~\eqref{eq:rho}, defined by the following differential equation,
\begin{equation}
    \label{eq:rho-pdeq}
	\dfrac{\partial_{\delta}\rho}{\partial_{R}\rho}=\dfrac{q_3}{q_0q_1}.
\end{equation}
	Then the Thiele equations can be splited into three blocks:
\begin{equation}
\label{eq:Thiele2}
    \begin{aligned}
	-\alpha q_2 \dot{a} + a \dot{\varphi} = \ell_w^2\partial_a\mathfrak{F},
	&\;
	-\dot{a}-\alpha q_2 a \dot{\varphi} 
    =0,\qquad
	\\
	-\alpha(\tilde{q}_1^2/q_1)  \dot{\rho}+\ell_w\tilde{q}_1 \dot{\eta}=\ell_w^2\partial_{\rho}\mathfrak{F},
	&\;
	-\ell_w^{-1}\tilde{q}_1 \dot{\rho}-\alpha q_1 \dot{\eta}=\partial_{\eta}\mathfrak{F},
	\\
	-\alpha (q_4-q_3^2/q_1)\dot{\delta}=\ell_w^2\partial_{\delta}\mathfrak{F},
	\end{aligned}
\end{equation}
	where time is scaled in  
	${t_0=M_s\ell_w^3/(2\gamma Ad_F)}$, and ${\tilde{q}_1={q}_0q_1/\partial_{R}\rho}$.

    Since dissipation is usually weak, $\alpha\ll1$, we can describe qualitatively the evolution of the skyrmion far away from the tip in the following way. Firstly, the variation rate~$\ell_w^{-1}\dot\delta\sim\alpha^{-1}$ is much larger than the other rates, $\ell_w^{-1}\dot{\bm{a}}$, $\ell_w^{-1}\dot{\rho}$, and $\dot{\eta}$. Physically this means that effective wall thickness~$\delta$ adapts much quicker than the other parameters and can be considered as some known function of $\bm{a}$, $\rho$, and $\eta$. 
    
    Secondly, distance~${a}$ changes much slower that the azimuthal angle~${\varphi}$ (and, obviously, $\rho$ and $\eta$), 
    \begin{eqnarray}
	\dot{a}\approx-\alpha q_2 \ell_w^2\partial_a\mathfrak{F},
    \qquad
    \dot{\varphi} \approx a^{-1}\partial_a\mathfrak{F},
	\end{eqnarray}
    therefore skyrmion moves along the spiral trajectory.

    Finally, the $\rho$ and $\eta$ oscillates with the same frequency (and dissipation proportional to $\alpha$) near its quasi-equilibrium values when assuming $a$ is almost constant. 

\section{Asymptotic expressions at $R\gg\ell_w$\label{app:asymp}}

Here we provide the details for the calculation of Eq.~\eqref{eq:F_R>>1} for the free energy~$\mathcal{F}$ at large skyrmion radius, $R\gg\ell_w$, and solve differential equation for the adjusted radius $\rho$, see Eq.~\eqref{eq:rho-pdeq}. 

\subsection{Expression for $q_i(R,\delta)$ and $p(R,\delta)$ without electric field} 

In this limiting case we assume that the skyrmion is a circle domain with radius~$R\gg\ell_w$ and wall of thickness $\delta\approx\ell_w$, and its magnetization can be described by skyrmion angle from Eq.~\eqref{eq:DW_R>>1}. Then we can calculate functions $q_i(R,\delta)$ from Eqs.~\eqref{eq:qi} and $p$ without electric field, $\beta=0$, asymptotically,
\begin{subequations}
    \label{eq:qi-asym}
	\begin{eqnarray}
		q_0&\approx&\dfrac{\ell_w}{\delta},
		\quad
		q_{1}\approx \dfrac{R\delta}{\ell_w^2},
		\quad
		q_{2}\approx\dfrac{R}{2\delta}+\dfrac{\delta}{2R},
		\\
		q_{3}&\approx&\dfrac{\pi^2\delta}{12\ell_w},
		\quad
        q_{4}\approx\dfrac{\pi^2 R}{12\delta},
        \quad
        p\big|_{\beta=0}=- \dfrac{\pi R}{\ell_w}\, \epsilon_0.
	\end{eqnarray}
	\end{subequations}
The calculation is straightforward and partly can be found in Ref.~\cite{Wang2018}. Note that above expressions for functions $q_0$, $q_1$, $q_3$, $q_4$, and $p\big|_{\beta=0}$ are calculated with exponential accuracy, i.e. neglecting the terms of order $\exp(-R/\ell_w)$, while function $q_2$ is chopped to the terms of order $(R/\ell_w)^{-3}$.

\subsection{Free energy}

Then to calculate the total free energy ${\mathcal{F}=\mathcal{F}_{\text{magn}}+\mathcal{F}_{\text{ME}}}$, see Eq.~\eqref{eq:F-via-q}, we should use expressions for $q_1$, $q_2$, and $p\big|_{\beta=0}$ to get first two terms and the first term inside square brackets in Eq.~\eqref{eq:F_R>>1}. The second term in square brackets is resulted from the ME energy and, in the main approximation in $R\gg \ell_w$, can be expressed as
\begin{equation}
	\beta\bar{\mathcal{E}}_{a}(R)f(R-a)=\dfrac{\beta}{\pi}
	\int_{0}^{\infty} \dfrac{dr}{\ell_w} \dfrac{\bar{\mathcal{E}}_{a}(r)}{\cosh[(r-R)/\ell_w]}.
	\label{eq:f_gen}
\end{equation}

To calculate the last integral approximately we note that function $\bar{\mathcal{E}}_{a}(r)$ has a Lorentzian peak near $r=a$, see $\bar{h}^2(r-a)$ in the denominator in Eq.~\eqref{eq:eps_a}, and slowly changes far away from that point. Analogously, function $1/\cosh[(r-R)/\ell_w]$ represents the peak near $r=R$. Then, we can approximate expression under integral by rational function,
\begin{equation}
	\dfrac{\bar{\mathcal{E}}_{a}(r)}{\cosh[(r-R)/\ell_w]}\approx
	\dfrac{\bar{\mathcal{E}}_{a}(R)}{\{[(r-R)/2\ell_w]^2+1\}^2}\dfrac{\bar{h}^2(R-a)}{\bar{h}^2(r-a)},
\end{equation}
that can be easily integrated and yields
\begin{equation}
f(z)=\frac{(h^2+z^2) \left[(h+4\ell_w) (h+2\ell_w)^2/h+z^2\right]}{\left[(h+2\ell_w)^2+z^2\right]^2}.
	\label{eq:f_simpl}
\end{equation}
Note that if distance~$a$ between skyrmion center and the charged tip significantly differs from skyrmion radius, $|z|=|R-a|\gg\ell_w$, then function $f(R-a)$ is equal to $1$. Then expression in square bracket in Eq.~\eqref{eq:F_R>>1} becomes $[\epsilon_0+\beta\bar{\mathcal{E}}_{a}(R)]$, which simply means that effective DMI should be taken at the skyrmion domain wall, $r=R$. However, when radius approximately equal to distance, $|R-a|\sim \ell_w\ll R$, e.g., skyrmion is attached to the tip, see subsection~\ref{sec:diagram}, function $f(R-a)$ deviates from $1$, in the following range: $(1+\ell_w/h)^{-1}<f(z)<1+\ell_w/h$. 

\subsection{Adjusted radius}

In the limit $R\gg\ell_w$ Eq.~\eqref{eq:rho-pdeq} takes the following form,
\begin{equation}
    \label{eq:rho-pdeq-asym}
	\dfrac{\partial_{\delta}\rho}{\partial_{R}\rho}=\dfrac{\pi^2\delta}{12R}.
\end{equation}
that can be solved explicitly,
\begin{equation}
    \label{eq:rho-pdeq-sol}
	\rho=F(R^2+\pi^2\delta^2/12),
\end{equation}
where $F$ is the arbitrary function, which can be naturally taken as the square root. Therefore, we get the adjusted radius~$\rho$ as
\begin{equation}
    \label{eq:rho}
	\rho=\sqrt{R^2+\dfrac{\pi^2\delta^2}{12}}\approx R+\dfrac{\pi^2\delta^2}{12R}.
\end{equation}

\section{Multiple solution of Euler-Lagrange equation with non-uniform electric field\label{sec:multi}}

\begin{figure*}[!t]
    \includegraphics[width=0.9\textwidth]{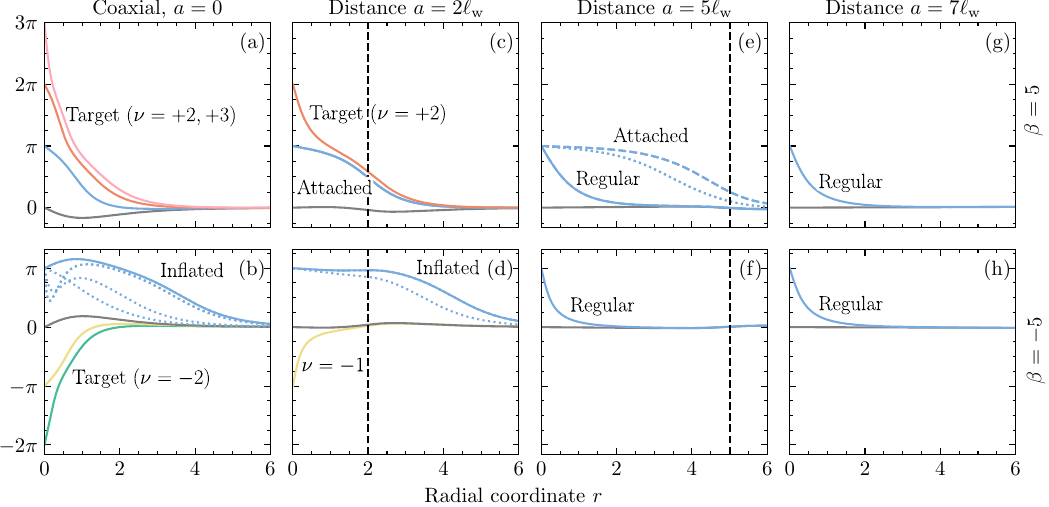}
    \caption{
		Example of solutions found by solving the EL equation~\eqref{eq:euler_lagrange} for different values of distance~$a$ and field intensities~$\beta$. 
		Gray lines represent solutions without a skyrmion
        ($\nu=0$). 
		Blue and yellow lines depict regular skyrmions of positive and negative chirality, $\nu=\pm1$, respectively. 
		Orange, green, and pink lines correspond to target skyrmions with $\nu=\pm2, +3$, respectively.  
		Multiple solutions of the same kind are distinguished by different line styles, where a solid line represents the solution with the lowest energy, dashed lines represent other stable solutions, and dotted lines represent unstable solutions.
		Black dashed vertical lines show position of the tip relative to the skyrmion center.
	}
    \label{fig:skyrmion-examples}
\end{figure*}

Figure~\ref{fig:skyrmion-examples} shows solutions of Eq.~\eqref{eq:euler_lagrange} with $\epsilon_{\rm eff}$ given by Eq.~\eqref{eq:eps_a} for certain values of the distance $a$, the field intensity $\beta$, and the acceptable number~$\nu$. 

Generally, the solutions of Euler-Lagrange (EL) equation can correspond not only to the minimum of the free energy, but to the saddle-like solutions. The former are stable (or quasi-stable) and the latter are unstable. Therefore, solutions shown in Fig.~\ref{fig:skyrmion-examples}  correspond to stable (solid or dashed line) or unstable (dotted line) states of skyrmion with $|\nu|=1$ (blue and yellow lines), skyrmionium  with $|\nu|=2$ (orange and green lines) or target skyrmion  with $|\nu|>2$ (rose lines). Also the solutions with $\nu=0$ describe the no-skyrmion states (gray lines).
The solid curves in Fig.~\ref{fig:skyrmion-examples} correspond to the lowest-energy stable configurations for given $\nu$, while 
dashed curves denote higher-energy metastable configurations. 
Hence, although multiple mathematical solutions of the EL equation exist for the same parameters, only a subset of them can be realized as stable skyrmion states.

The number of available solutions changes significantly with variation of the distance $a$ and the field intensity $\beta$. 
For small distances $a$, where the influence of the field is strongest, the electric field significantly 
leads to the appearance of multiple stable and unstable solutions.
In particular, for sufficiently large positive or negative values of $\beta$, several target skyrmion solutions with different 
numbers $\nu$
may coexist with regular solutions, cf. panels (a) and (b) in Fig.~\ref{fig:skyrmion-examples}. 
As the distance $a$ between skyrmion and the charged tip increases, the influence of the field weakens, reducing the number of allowed configurations, cf. panels (c)-(f) in Fig.~\ref{fig:skyrmion-examples}. 
At sufficiantly large $a$, only regular skyrmions and the no-skyrmion state remain stable, cf. panels (g) and (h) in Fig.~\ref{fig:skyrmion-examples}.  

One can compare Figs.~\ref{fig:diagram} and~\ref{fig:skyrmion-examples} and identify correspondent regimes, labeled as regular, attached, and inflated skyrmions, as well as target skyrmions.

\section{Micromagnetic framework\label{app:MF}}

In this section, we describe the numerical approach of the micromagnetic simulations used in this work. 

As mentioned in the main text, the time evolution of the magnetization 
is modeled using the Landau--Lifshitz--Gilbert (LLG) equation~\eqref{eq:LLG}, while 
the stationary magnetization is obtained
by minimizing the total free energy $\mathcal{F}=\mathcal{F}_{\text{magn}}+\mathcal{F}_{\text{ME}}$, see Eqs.~\eqref{eq:MagFe} and~\eqref{eq:w_cnv}. 

The micromagnetic simulations (MMS) are a numerical approach in which each group of close spins is treated as a single macrospin, represented by a classical magnetic vector placed in a grid cell with periodic boundary condition, for the overall configuration of which we seek the configuration that minimizes the total energy.
This macrospin approximation is possible due to the existence of exchange interaction between close spins, constraining their relative orientation to be almost parallel.

In order to find the minimum-energy configuration, two numerical methods can be employed -- an energy driver and a time driver. The energy driver uses gradient descent of the energy in order to evolve the system towards a state with lower energy. After a sufficient number of steps, the system converges to a state with locally minimal energy. Different starting configurations can be used in order to explore multiple possible minima.   

The time driver uses the LLG equation to evolve the system in time towards the state with minimal energy due to the existence of a dissipative term, which reduces the overall energy with each time step. This approach is more precise and realistic, showing the spiraling motion of the skyrmion near the tip, although this approach is more computationally expensive than the energy driver, which seeks the minimum directly. 
In the cases described in this paper, both methods show agreement on the final result of the modeling, although proper skyrmion motion during its evolution is only obtained with the time driver. 

The simulations were performed using the \textit{Object-Oriented MicroMagnetic Framework} (OOMMF)~\cite{OOMMF} within the \textit{Ubermag} environment~\cite{Ubermag}. The computational domain was chosen as a square of size ${2 \times 2~\mu\mathrm{m}^2}$, discretized into a uniform grid with a cell size of ${2.5~\mathrm{nm}}$.  The chosen discretization ensures sufficient spatial resolution to accurately capture the structure of skyrmions and their domain walls. The damping parameter ${\alpha = 0.05}$ was selected to provide efficient convergence to stationary states.

\bibliography{skyrmions}

\end{document}